\PassOptionsToPackage{unicode}{hyperref}
\PassOptionsToPackage{hyphens}{url}
\PassOptionsToPackage{dvipsnames,svgnames,x11names}{xcolor}
\documentclass[
  12pt,
  11pt]{article}

\usepackage{amsmath,amssymb}
\usepackage{setspace}
\usepackage{iftex}
\ifPDFTeX
  \usepackage[T1]{fontenc}
  \usepackage[utf8]{inputenc}
  \usepackage{textcomp} 
\else 
  \usepackage{unicode-math}
  \defaultfontfeatures{Scale=MatchLowercase}
  \defaultfontfeatures[\rmfamily]{Ligatures=TeX,Scale=1}
\fi
\usepackage{lmodern}
\ifPDFTeX\else  
\fi
\IfFileExists{upquote.sty}{\usepackage{upquote}}{}
\IfFileExists{microtype.sty}{
  \usepackage[]{microtype}
  \UseMicrotypeSet[protrusion]{basicmath} 
}{}
\makeatletter
\@ifundefined{KOMAClassName}{
  \IfFileExists{parskip.sty}{%
    \usepackage{parskip}
  }{
    \setlength{\parindent}{0pt}
    \setlength{\parskip}{6pt plus 2pt minus 1pt}}
}{
  \KOMAoptions{parskip=half}}
\makeatother
\usepackage{xcolor}
\ifx\paragraph\undefined\else
  \let\oldparagraph\paragraph
  \renewcommand{\paragraph}[1]{\oldparagraph{#1}\mbox{}}
\fi
\ifx\subparagraph\undefined\else
  \let\oldsubparagraph\subparagraph
  \renewcommand{\subparagraph}[1]{\oldsubparagraph{#1}\mbox{}}
\fi

\usepackage{longtable,booktabs,array}
\usepackage{calc} 
\usepackage{etoolbox}
\makeatletter
\patchcmd\longtable{\par}{\if@noskipsec\mbox{}\fi\par}{}{}
\makeatother
\IfFileExists{footnotehyper.sty}{\usepackage{footnotehyper}}{\usepackage{footnote}}
\makesavenoteenv{longtable}
\usepackage{graphicx}
\makeatletter
\def\maxwidth{\ifdim\Gin@nat@width>\linewidth\linewidth\else\Gin@nat@width\fi}
\def\maxheight{\ifdim\Gin@nat@height>\textheight\textheight\else\Gin@nat@height\fi}
\makeatother
\setkeys{Gin}{width=\maxwidth,height=\maxheight,keepaspectratio}
\makeatletter
\def\fps@figure{htbp}
\makeatother

\usepackage{dcolumn}
\makeatletter
\@ifpackageloaded{caption}{}{\usepackage{caption}}
\AtBeginDocument{%
\ifdefined\contentsname
  \renewcommand*\contentsname{Table of contents}
\else
  \newcommand\contentsname{Table of contents}
\fi
\ifdefined\listfigurename
  \renewcommand*\listfigurename{List of Figures}
\else
  \newcommand\listfigurename{List of Figures}
\fi
\ifdefined\listtablename
  \renewcommand*\listtablename{List of Tables}
\else
  \newcommand\listtablename{List of Tables}
\fi
\ifdefined\figurename
  \renewcommand*\figurename{Figure}
\else
  \newcommand\figurename{Figure}
\fi
\ifdefined\tablename
  \renewcommand*\tablename{Table}
\else
  \newcommand\tablename{Table}
\fi
}
\@ifpackageloaded{float}{}{\usepackage{float}}
\floatstyle{ruled}
\@ifundefined{c@chapter}{\newfloat{codelisting}{h}{lop}}{\newfloat{codelisting}{h}{lop}[chapter]}
\floatname{codelisting}{Listing}

\makeatother
\makeatletter
\@ifpackageloaded{caption}{}{\usepackage{caption}}
\@ifpackageloaded{subcaption}{}{\usepackage{subcaption}}
\makeatother
\makeatletter
\@ifpackageloaded{tcolorbox}{}{\usepackage[skins,breakable]{tcolorbox}}
\makeatother
\makeatletter
\@ifundefined{shadecolor}{\definecolor{shadecolor}{rgb}{.97, .97, .97}}
\makeatother
\ifLuaTeX
  \usepackage{selnolig}  
\fi
\usepackage[]{natbib}
\IfFileExists{bookmark.sty}{\usepackage{bookmark}}{\usepackage{hyperref}}
\IfFileExists{xurl.sty}{\usepackage{xurl}}{} 
\hypersetup{
  pdftitle={Spatial Data Analysis},
  pdfauthor={Tobias Rüttenauer},
  pdfkeywords={Spatial Econometrics, Spatial Data Analysis, Spatial
Dependence, Spatial Spillovers, London House Prices, Handbook},
  colorlinks=true,
  linkcolor={blue},
  filecolor={Maroon},
  citecolor={Blue},
  urlcolor={Blue},
  pdfcreator={LaTeX via pandoc}}

\begin{document}

\def\spacingset#1{\renewcommand{\baselinestretch}%
{#1}\small\normalsize} \spacingset{1}


\date{March 18, 2025}
\title{\bf Spatial Data Analysis}
\author{
Tobias Rüttenauer\thanks{I would like to thank Robert Vief for valuable
comments on an earlier version of this manuscript. I am also grateful to
workshop participants for their comments on the respective teaching
materials.}\\
UCL Social Research Institute, University College London\\
}
\maketitle

\bigskip
\bigskip
\begin{abstract}
This handbook chapter provides an essential introduction to the field of
spatial econometrics, offering a comprehensive overview of techniques
and methodologies for analysing spatial data in the social sciences.
Spatial econometrics addresses the unique challenges posed by spatially
dependent observations, where spatial relationships among data points
can be of substantive interest or can significantly impact statistical
analyses. The chapter begins by exploring the fundamental concepts of
spatial dependence and spatial autocorrelation, and highlighting their
implications for traditional econometric models. It then introduces a
range of spatial econometric models, particularly spatial lag, spatial
error, spatial lag of X, and spatial Durbin models, illustrating how
these models accommodate spatial relationships and yield accurate and
insightful results about the underlying spatial processes. The chapter
provides an intuitive guide on how to interpret those different models.
A practical example on London house prices demonstrates the application
of spatial econometrics, emphasising its relevance in uncovering hidden
spatial patterns, addressing endogeneity, and providing robust estimates
in the presence of spatial dependence.
\end{abstract}

\noindent%
{\it Keywords:} Spatial Econometrics, Spatial Data Analysis, Spatial
Dependence, Spatial Spillovers, London House Prices, Handbook
\vfill

\newpage
\spacingset{1.9} 
\ifdefined\Shaded\renewenvironment{Shaded}{\begin{tcolorbox}[boxrule=0pt, sharp corners, enhanced, interior hidden, frame hidden, borderline west={3pt}{0pt}{shadecolor}, breakable]}{\end{tcolorbox}}\fi

\setstretch{1.25}
\newcommand{\Exp}{\mathrm{E}}
\newcommand\given[1][]{\:#1\vert\:}
\newcommand{\Cov}{\mathrm{Cov}}
\newcommand{\Var}{\mathrm{Var}}
\newcommand{\rank}{\mathrm{rank}}
\newcommand{\bm}[1]{\boldsymbol{\mathbf{#1}}}
\newcommand{\tr}{\mathrm{tr}}
\newcommand{\plim}{\operatornamewithlimits{plim}}
\newcommand{\diag}{\mathrm{diag}}

\hypertarget{introduction}{%
\section{Introduction}\label{introduction}}

The availability of spatial data for social sciences has rapidly
increased over the past decade. Various spatial packages have been
implemented in standard statistical software and have steadily been
updated \citep{Bivand.2022} in addition to already existing software in
Geo-Information Systems (GIS) such as ArcGIS or QGIS. At the same time,
many empirical social science papers investigate research questions with
an explicit spatial focus. Examples of spatial topics in the social
sciences include labour market dynamics
\citep{Marten.2019, Nisic.2017, Zoch.2021}, processes of residential
segregation \citep{Roberto.2018, Toth.2021} and gentrification
\citep{Fransham.2020, Zapatka.2021}, the spatial distribution of
environmental goods or bads
\citep{Boillat.2022, Junger.2022, Ruttenauer.2018a}, the consequences of
extreme weather events
\citep{Ogunbode.2019, Hoffmann.2022, Ruttenauer.2023a} or the access to
infrastructural conditions
\citep{Moreno-Monroy.2018, Liao.2020, Wiedner.2022}.

In general, spatial data is structured like conventional data (e.g.~a
dataset with variables), but has one additional dimension: every
observation is linked to some geo-spatial information. Most common types
of spatial information are points, lines, or polygons (vector data) or
raster data. Similar to the time dimension in panel data, this adds an
additional layer of information and connectivity between units. As with
panel data, we could thus proceed as if we had conventional data and
ignore the spatial dimension. This comes however with two distinct
problems. First, we waste potentially interesting information, that may
help us to understand the underlying social processes. Second, we will
end up with biased inferential statistics and biased point estimates in
some cases if we ignore the underlying spatial dependence.

There are various techniques to model spatial dependence and spatial
processes \citep{LeSage.2009}. Here, we will cover the most common
spatial econometric models. Generally, spatial regression models make
some assumptions about the source of spatial dependence observed in the
data and then account for this dependence in the specified model. What
makes spatial regression models more complicated than panel models is
the ambiguous direction and circular nature of the dependence. I may
influence my neighbour, but my neighbour may also influence me
(interdependence). Moreover, if someone influences a third neighbour,
they may be neighbours of my neighbour -- 2nd order neighbour of me --
which will then influence me as well (diffusion). However, we may also
not influence each other at all but just be affected by the same
exogenous shock, thus making our observed values more similar (common
confounding).

The chapter proceeds as follows. First, we will briefly introduce the
concept of spatial connectivity \(\boldsymbol{\mathbf{W}}\) and spatial
dependence and clarify why conventional regression techniques may fail
with spatial dependence. We will then provide an overview of the most
common spatial regression models. In a further step, we will demonstrate
how to interpret summary measures of the coefficients of these models,
which becomes more complicated in the case of spatial interdependence.
Lastly, we use the relation between neighbourhood characteristics and
house prices in London as an applied example.

\hypertarget{spatial-weights}{%
\section{Spatial weights}\label{spatial-weights}}

Given the geographical information of spatial data (i.e.~the location of
each unit), we can form relationships between units: which units are
closer or further away from each other. Similar to network analysis, we
have to set up a measure that defines which units are connected to each
other and how they are connected (e.g.~the magnitude of connectivity).
There are some obvious measures that can be used to define these
relations with spatial data: adjacency and proximity.

The connectivity between units is usually represented in a matrix
denoted \(\boldsymbol{\mathbf{W}}\). The spatial weights matrix
\(\boldsymbol{\mathbf{W}}\) is an \(N \times N\) dimensional matrix,
where each element \(w_{ij}\) of this matrix specifies the relation or
connectivity between each pair of units \(i\) and \(j\).

\begin{equation} 
        \boldsymbol{\mathbf{W}} = 
        \begin{bmatrix} 
            w_{11} & w_{12} & w_{13} & \dots & w_{1n} \\
            w_{21} & w_{22} & w_{23} & \dots & w_{2n} \\
            w_{31} & w_{32} & w_{33} & \dots & \vdots \\
            \vdots & \vdots & \vdots & \ddots & \vdots \\
            w_{n1} & w_{n2} & w_{n3} & \dots     & w_{nn} 
            \end{bmatrix}
\end{equation}

In the example above, \(w_{31}\) describes the relationship between unit
3 and unit 1, while \(w_{2n}\) describes how unit 2 and unit \(n\) are
connected. The diagonal elements
\(w_{i,i}= w_{1,1}, w_{2,2}, \dots, w_{n,n}\) of
\(\boldsymbol{\mathbf{W}}\) are always zero: no unit is a neighbour of
itself. This is not true for spatial multiplier matrices (as we will see
later). Contiguity weights are a very common type of spatial weights.
This is a binary specification, taking the value 1 for neighbouring
units (queens: sharing a common edge; rook: sharing a common border),
and 0 otherwise. See for instance \citet{Pebesma.2023} for more detailed
information about spatial relations.

Contiguity weights matrices are usually sparse matrices and keep
relations relatively simple and easy to interpret. However, they often
create island, i.e.~units without any neighbours, which can be
problematic for spatial regression models. Another common type of
connectivity measures is distance based weights. For instance, inverse
distance weights assign higher weights to more proximate units
\(w_{i,j} = \frac{1}{d_{ij}^\alpha}\), where distance is usually
discounted by a spatial decay factor \(\alpha\). It is often recommended
to specify a distance threshold (e.g.~100km) to get rid of very small
non-zero weights for very distant units. There is an ongoing debate
about the importance of spatial weights for spatial econometrics and
about the right way of specifying weights matrices
\citep{LeSage.2014a, Neumayer.2016}.

\hypertarget{normalization}{%
\subsection{Normalization}\label{normalization}}

Normalizing ensures that the parameter space of the spatial multiplier
in regression models is restricted to \(-1 < \rho > 1\), and the
multiplier matrix is non-singular (more on this later). The important
message is: normalizing the weights matrix is always a good idea.
Otherwise, the spatial parameters may blow up -- if they can be
estimated at all. Normalising also ensures an easy interpretation of
spillover effects (as we see later). Again, how to normalize a weights
matrix is subject of debate \citep{LeSage.2014a, Neumayer.2016}.

Row-normalization divides each non-zero weight by the sum of all weights
of unit \(i\), which is the sum of the row \(i\):
\(\frac{w_{ij}}{\sum_j^n w_{ij}}\). With contiguity weights and
row-normalisation, spatially lagged variables contain the mean of the
respective variable among the neighbours of \(i\). However, proportions
between units such as distances get lost due to row-normalisation, which
can be problematic if one is theoretically interested in using
inverse-distance based weights. It also induces asymmetries, as
different units have different numbers of neighbours:
\(w_{ij} \neq w_{ji}\).

Another common way of standardization is maximum eigenvalues
normalization. Maximum eigenvalues normalization divides each non-zero
weight by the overall maximum eigenvalue of the entire matrix
\(\lambda_{max}\): \(\frac{\boldsymbol{\mathbf{W}}}{\lambda_{max}}\).
Each element of \(\boldsymbol{\mathbf{W}}\) is divided by the same
scalar value, which preserves the relations. It keeps proportions of
connectivity strengths across rows, which is relevant for distance based
\(\boldsymbol{\mathbf{W}}\). I thus recommend maximum eigenvalues
normalization for distance based neighbours weights. However,
interpretation may become more complicated.

\hypertarget{spatial-dependence}{%
\subsection{Spatial dependence}\label{spatial-dependence}}

`Everything is related to everything else, but near things are more
related than distant things' \citep{Tobler.1970}. Tobler's first law of
geography has been used extensively (13,690 citation in 2025-03) to
describe spatial dependence. In practical term, this means that close
observations are more likely to exhibit similar values on some of their
characteristics, and we cannot handle observations as if they were
independent.

\begin{figure}

{\centering \includegraphics{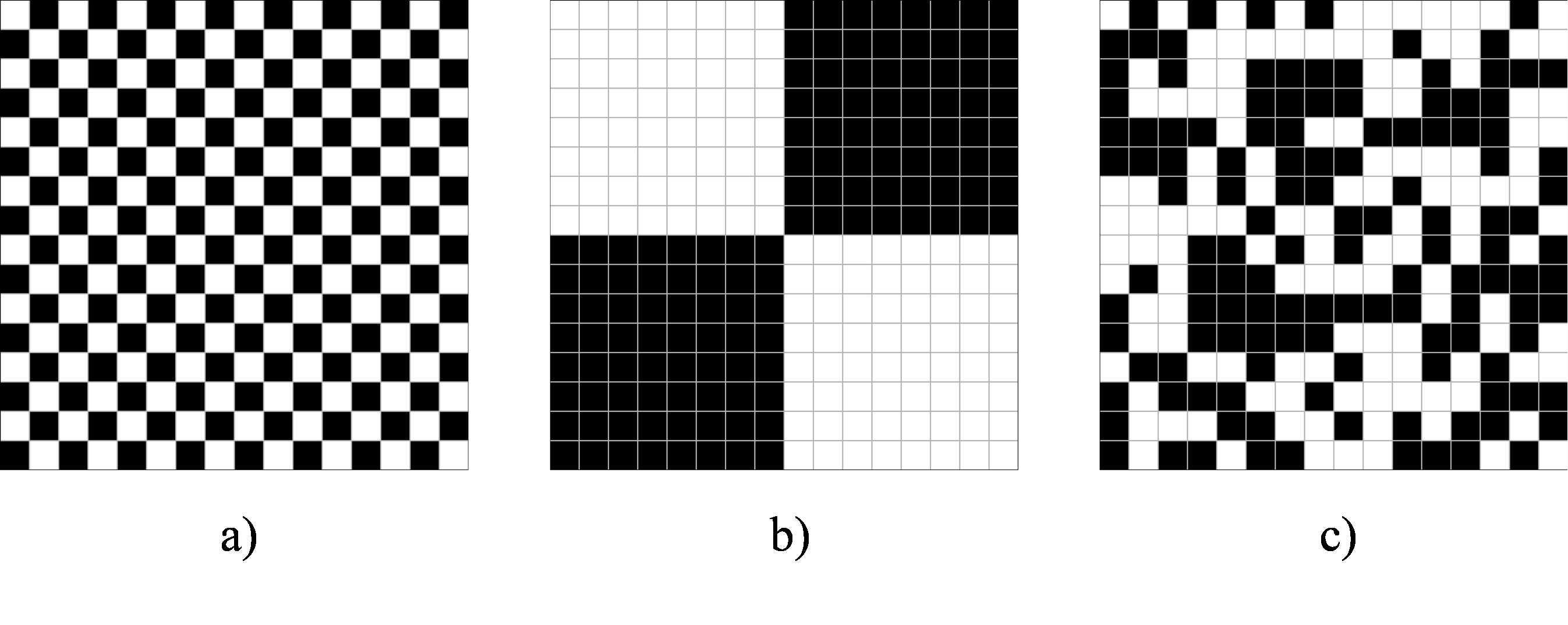}

}

\caption{\label{fig-chess}Forms of spatial dependence: a) perfect
negative autocorrelation, b) nearly perfect positive autocorrelation, c)
random.}

\end{figure}

There is a very easy and intuitive way of detecting spatial
autocorrelation: look at the map. Below we can see three distinct
patterns. Figure Figure~\ref{fig-chess} a) has perfect negative
auto-correlation. Every black unit is surrounded by white units, and
every white unit is surrounded by black units. Figure
Figure~\ref{fig-chess} b) has very strong positive autocorrelation. Most
white units are surrounded only by white units, and most black units are
surrounded by only black units. Figure Figure~\ref{fig-chess} c), by
contrast, is generated by a random process, although even here one is
inclined to observe some degree of clustering.

Would our interpretation be the same if we aggregate the data to four
larger areas / districts using the average within each of the four
districts? We would actually draw very different conclusions. It is thus
important to keep in mind that spatial dependence is a also a result of
spatial boundaries and potential higher-level processes generating an
outcome \citep{Wong.2009}. If a variable was measured on the district
level and we assign those district-level measures to the lower
neighbourhood level, we will artificially introduce spatial dependence /
clustering in our data.

Given our spatial data, we can use various statistical measures to test
whether there is spatial dependence. The most common statistic for
spatial dependence or autocorrelation is Moran's I, which goes back to
\citet{Moran.1950} and \citet{Cliff.1972}. For more extensive materials
on Moran's I, see for instance \citet{Kelejian.2017}, Chapter 11. We
first define a neighbours weights matrix \(\boldsymbol{\mathbf{W}}\),
and the Global Moran's I test statistic is calculated as

\begin{equation} 
        \boldsymbol{\mathbf{I}}  = \frac{N}{S_0}  
        \frac{\sum_i\sum_j w_{ij}(y_i-\bar{y})(y_j-\bar{y})}
            {\sum_i (y_i-\bar{y})^2}, \text{where } S_0 = \sum_{i=1}^N\sum_{j=1}^N w_{ij}
\end{equation}

In the case of row-standardized weights, \(S_0 = N\). Moran's \(I\)
measures the correlation between neighbouring values: how does my
\(y_i\) correlate with the average \(y_j\) of my neighbours? Negative
values indicate negative autocorrelation, values around zero (not zero
exactly) indicate no autocorrelation, and positive values indicate
positive autocorrelation. Moran's \(I\) can also be calculated for the
residuals from an estimated model (e.g.~non-spatial OLS), which allows
to test for remaining autocorrelation after accounting for potential
confounders. Note that local indicators of spatial autoccorelation
(LISA) and clustering -- such as local Moran's I or Geary's C -- can be
a relevant method of analysis by itself
\citep{Anselin.1995, Pebesma.2023}.

\hypertarget{bias-in-non-spatial-ols}{%
\section{Bias in non-spatial OLS}\label{bias-in-non-spatial-ols}}

So, why should we care about spatial dependence? First, spatial
dependence violates standard assumptions of common non-spatial
estimators. Second, spatial dependence itself can provide important
information about the social processes that generated the data we
observe.

Let us start with a linear model in the non-spatial setting. Here,
\(\boldsymbol{\mathbf{y}}\) is the outcome or dependent variable
(\(N \times 1\)), \(\boldsymbol{\mathbf{X}}\) are various exogenous
covariates (\(N \times k\)), and \(\boldsymbol{\varepsilon}\)
(\(N \times 1\)) is the error term. We are usually interested in an
estimate for the \(k \times 1\) coefficient vector
\(\boldsymbol{\beta}\).

\[
{\boldsymbol{\mathbf{y}}}={\boldsymbol{\mathbf{X}}}{\boldsymbol{\beta}}+ {\boldsymbol{\varepsilon}}
\]

The work-horse for estimating \(\boldsymbol{\beta}\) in the social
science is the OLS estimator \citep{Wooldridge.2010}, which is given by
the form:

\[
\hat{\beta}=({\boldsymbol{\mathbf{X}}}^\intercal{\boldsymbol{\mathbf{X}}})^{-1}{\boldsymbol{\mathbf{X}}}^\intercal{\boldsymbol{\mathbf{y}}}.
\]

This OLS estimator hinges on a few assumptions, among them that the
underlying sample observations are independent and identically
distributed (i.i.d). This assumption is often violated with spatial
data. Another (more important) assumptions is the absence of any omitted
(residual) variables that are related to \(\boldsymbol{\mathbf{Y}}\) and
\(\boldsymbol{\mathbf{X}}\):
\(\mathrm{E}(\epsilon_i|\boldsymbol{\mathbf{X}}_i) = 0\). This
assumption is violated when our neighbours' characteristics influence
our covariates and our outcome.

So, does spatial dependence allways induce bias in non-spatial
estimators? No, the best answer is: \emph{it depends}
\citep{Betz.2020, Cook.2020, Pace.2010, Ruttenauer.2022a}. The easiest
way to think of it is analogous to the well-kown omitted variable bias
\citep{Betz.2020, Cook.2020}:

\[
plim~\hat{\beta}_{OLS}= \beta  + \gamma \frac{\mathrm{Cov}(\boldsymbol{\mathbf{x}}, \boldsymbol{\mathbf{z}})}{\mathrm{Var}(\boldsymbol{\mathbf{x}})},
\]

where \(\boldsymbol{\mathbf{z}}\) is some omitted variable, and
\(\gamma\) is the conditional effect of \(\boldsymbol{\mathbf{z}}\) on
\(\boldsymbol{\mathbf{y}}\). Now imagine that the neighbouring values of
the dependent variable
\(\boldsymbol{\mathbf{W}} \boldsymbol{\mathbf{y}}\) are autocorrelated
to the focal unit's outcome. We denote this correlation with
\(\rho > 0\). Imagine further that the covariance between the focal
unit's exogenous covariate \(\boldsymbol{\mathbf{x}}\) and my
neighbour's outcome \(\boldsymbol{\mathbf{W}} \boldsymbol{\mathbf{y}}\)
is not zero (my covariate affects my neighbours' outcome). Then we will
have an omitted variable bias due to spatial dependence:

\[
plim~\hat{\beta}_{OLS}= \beta  + \rho \frac{\mathrm{Cov}(\boldsymbol{\mathbf{x}}, \boldsymbol{\mathbf{W}} \boldsymbol{\mathbf{y}})}{\mathrm{Var}(\boldsymbol{\mathbf{x}})} \neq \beta,
\]

\hypertarget{spatial-regression-models}{%
\section{Spatial Regression Models}\label{spatial-regression-models}}

\begin{figure}

{\centering \includegraphics{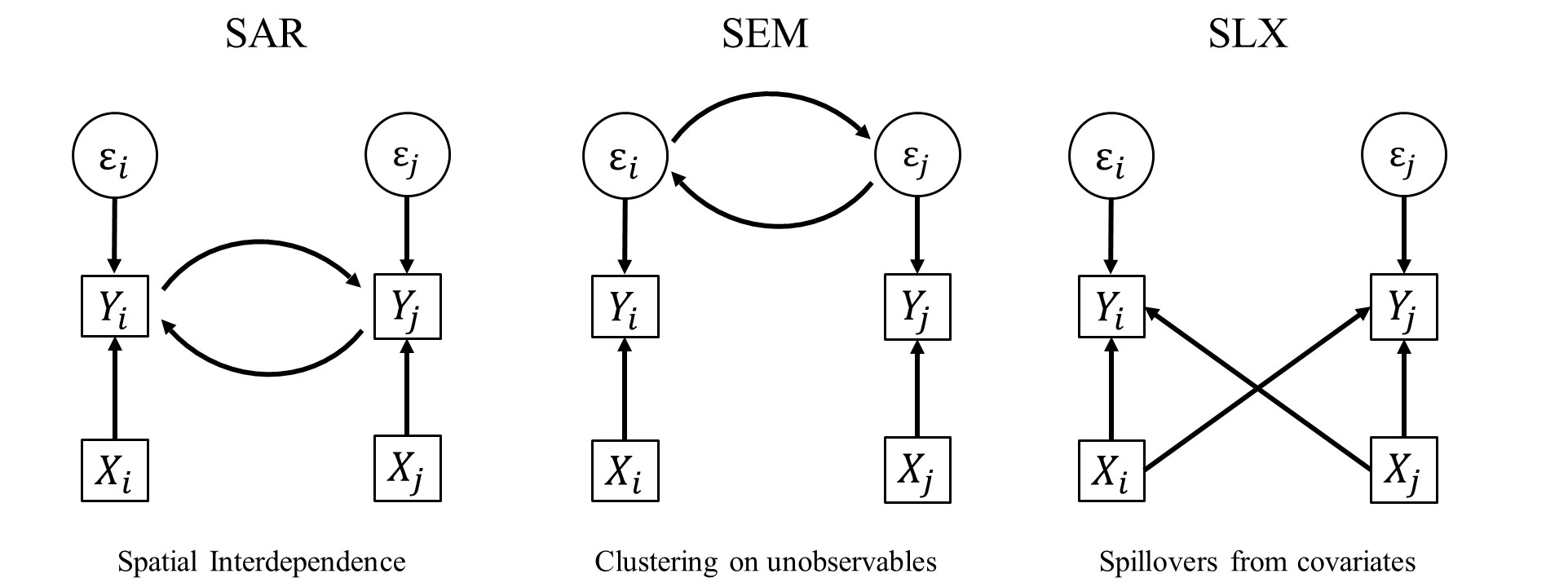}

}

\caption{\label{fig-models}Spatial regression models and their
assumptions about spatial dependence.}

\end{figure}

Spatial regression models do not only overcome the potential bias, they
also help us to understand the spatial processes happening in the
underlying data. Broadly, spatial dependence in some characteristics can
be the result of three different processes: a) Spatial interdependence,
b) Clustering in unobservables, and c) Spillovers from covariates. As
shown in Figure Figure~\ref{fig-models}, there are three basic ways of
incorporating spatial dependence: the Spatial Autoregressive Model (SAR)
accounts for spatial interdependence, the Spatial Error Model (SEM) for
clustering on unobservables, and the Spatially lagged X Model (SLX) for
spillovers from covariates. Moreover, they can be further combined. As
before, the \(N \times N\) spatial weights matrix
\(\boldsymbol{\mathbf{W}}\) defines the spatial relationship between
units.

\hypertarget{spatial-autoregressive-model-sar}{%
\subsection{Spatial Autoregressive Model
(SAR)}\label{spatial-autoregressive-model-sar}}

The Spatial Autoregressive Model (SAR) model is by far the most
prominent spatial specification. It assumes spatial interdependence in
the outcome \(\boldsymbol{\mathbf{Y}}\) and incorporates this
interdependence in the model:

\begin{equation} 
        {\boldsymbol{\mathbf{y}}}=\alpha{\boldsymbol{\iota}}+\rho{\boldsymbol{\mathbf{W}}}{\boldsymbol{\mathbf{y}}}+{\boldsymbol{\mathbf{X}}}{\boldsymbol{\beta}}+ {\boldsymbol{\varepsilon}}
\end{equation}

Here, \(\rho\) denotes the strength of the spatial correlation in the
dependent variable (spatial autocorrelation): \emph{your outcome
influences my outcome} (\(> 0\): positive spatial dependence, \(< 0\):
negative spatial dependence, \(= 0\): traditional OLS model). Given that
we have normalised the weights matrix, \(\rho\) is defined in the range
of \([-1, +1]\).

\hypertarget{spatial-error-model-sem}{%
\subsection{Spatial Error Model (SEM)}\label{spatial-error-model-sem}}

A second, also very common spatial model is the Spatial Error Model
(SEM). It assumes Clustering on unobservables, and thus models spatial
interdependence in the error term:

\begin{equation} 
\begin{split}
        {\boldsymbol{\mathbf{y}}}&=\alpha{\boldsymbol{\iota}}+{\boldsymbol{\mathbf{X}}}{\boldsymbol{\beta}}+{\boldsymbol{\mathbf{u}}},\\
        {\boldsymbol{\mathbf{u}}}&=\lambda{\boldsymbol{\mathbf{W}}}{\boldsymbol{\mathbf{u}}}+{\boldsymbol{\varepsilon}}
\end{split} 
\end{equation}

In this case, \(\lambda\) denotes the strength of the spatial
correlation in the errors of the model: \emph{your error influences my
errors} (\(> 0\): positive error dependence, \(< 0\): negative error
dependence, \(= 0\): traditional OLS model). Again, \(\lambda\) is
defined in the range of \([-1, +1]\).

\hypertarget{spatially-lagged-x-model-slx}{%
\subsection{Spatially lagged X Model
(SLX)}\label{spatially-lagged-x-model-slx}}

A third spatial model is called Spatially lagged X Model (SLX). It
assumes spillovers in the covariates. It specifies a relationship
between the covariate values of neighbours and the outcome of the focal
unit:

\begin{equation}
        {\boldsymbol{\mathbf{y}}}=\alpha{\boldsymbol{\iota}}+{\boldsymbol{\mathbf{X}}}{\boldsymbol{\beta}}+{\boldsymbol{\mathbf{W}}}{\boldsymbol{\mathbf{X}}}{\boldsymbol{\theta}}+ {\boldsymbol{\varepsilon}}
\end{equation}

In the SLX, \(\theta\) denotes the strength of the spatial spillover
effects from covariate(s) on the dependent variable: \emph{your
covariates influence my outcome}. In contrast to the previous two
specifications, \(\theta\) is defined like any other coefficient from a
conventional covariate. It is thus not bound to any range, and its scale
depends on the scale of the covariates in \(\boldsymbol{\mathbf{X}}\).

The dependence structure assumed in SAR and SEM has a circular element
(see Figure Figure~\ref{fig-models}). In A SAR model, my outcome
influences my neighbours' outcome, which then again influences my
outcome. In A SEM model, my error term influences my neighbours' error
term, which then again influences my error term. This also means that
SAR and SEM models cannot be estimated by conventional OLS estimators,
as they would suffer from simultaneity bias in the spatial
autoregressive term:

\begin{equation} 
\begin{split}
    \hat{\rho}_{OLS} &= \rho + \left[\left(\boldsymbol{\mathbf{W}}\boldsymbol{\mathbf{y}}     \right)^\top\left(\boldsymbol{\mathbf{W}}\boldsymbol{\mathbf{y}} \right)\right]^{-1}\left(\boldsymbol{\mathbf{W}}\boldsymbol{\mathbf{y}}    \right)^\top\varepsilon \\
    &= \rho + \left(\sum_{i = 1}^n \boldsymbol{\mathbf{y}}_{Li}^2\right)^{-1}\left(\sum_{i =     1}^{n}\boldsymbol{\mathbf{y}}_{Li}\epsilon_i\right),
\end{split}
\end{equation}

with \(\boldsymbol{\mathbf{y}}_{Li}\) defined as the \(i\)th element of
the spatial lag operator
\(\boldsymbol{\mathbf{W}}\boldsymbol{\mathbf{y}} = \boldsymbol{\mathbf{y}}_L\).
It can further be shown that the second part of the equation \(\neq 0\),
which demonstrates that OLS would provide a biased estimate of \(\rho\)
\citep{Franzese.2007, Sarrias.2023}.

A potential way of estimating SAR-like models is an instrumental
variable approach with 2SLS, where the autoregressive term is
instrumented by spatial lags of
\(\boldsymbol{\mathbf{H}} = \boldsymbol{\mathbf{X}}, \boldsymbol{\mathbf{W}}\boldsymbol{\mathbf{X}}, \boldsymbol{\mathbf{W}}^2\boldsymbol{\mathbf{X}}, ... , \boldsymbol{\mathbf{W}}^l\boldsymbol{\mathbf{X}},\)
\citep{Kelejian.1998}. SEM-like models can be estimated using
Generalized Method of Moments \citep{Kelejian.1999}. However, given the
improvements in computational power, it is now common to rely on Maximum
Likelihood estimation of spatial models \citep{Ord.1975, Anselin.1988}.
They start with some auxiliary regression to obtain initial estimates,
and then update them in further steps. For more details see
\citet{Bivand.2015}, \citet{LeSage.2009}, and \citet{Sarrias.2023}. The
R package \texttt{spatialreg}
\citep{Bivand.2015, Bivand.2021, Pebesma.2023} provides a series of
functions to calculate the ML estimators for all spatial models
considered here.

Moreover, there are models combining two sets of the above
specifications.

\hypertarget{spatial-durbin-model-sdm}{%
\subsection{Spatial Durbin Model (SDM)}\label{spatial-durbin-model-sdm}}

The spatial Spatial Durbin Model (SDM) integrates spatial
interdependence in the outcome and spatial spillovers in covariates by
combining SAR and SLX:

\begin{equation}
        {\boldsymbol{\mathbf{y}}}=\alpha{\boldsymbol{\iota}}+\rho{\boldsymbol{\mathbf{W}}}{\boldsymbol{\mathbf{y}}}+{\boldsymbol{\mathbf{X}}}{\boldsymbol{\beta}}+{\boldsymbol{\mathbf{W}}}{\boldsymbol{\mathbf{X}}}{\boldsymbol{\theta}}+ {\boldsymbol{\varepsilon}}
\end{equation}

\hypertarget{spatial-durbin-error-model-sdem}{%
\subsection{Spatial Durbin Error Model
(SDEM)}\label{spatial-durbin-error-model-sdem}}

The Spatial Durbin Error Model (SDEM) model integrates clustering on
unobservables and spillovers in covariates by combining SEM and SLX:

\begin{equation}
\begin{split}
        {\boldsymbol{\mathbf{y}}}&=\alpha{\boldsymbol{\iota}}+{\boldsymbol{\mathbf{X}}}{\boldsymbol{\beta}}+{\boldsymbol{\mathbf{W}}}{\boldsymbol{\mathbf{X}}}{\boldsymbol{\theta}}+ {\boldsymbol{\mathbf{u}}},\\
        {\boldsymbol{\mathbf{u}}}&=\lambda{\boldsymbol{\mathbf{W}}}{\boldsymbol{\mathbf{u}}}+{\boldsymbol{\varepsilon}}
\end{split}
\end{equation}

\hypertarget{combined-spatial-autocorrelation-model-sac}{%
\subsection{Combined Spatial Autocorrelation Model
(SAC)}\label{combined-spatial-autocorrelation-model-sac}}

The Combined Spatial Autocorrelation Model (SAC) assumes spatial
interdependence in the outcome and clustering on unobservables to be
present at the same time. It combines SAR and SEM:

\begin{equation}
\begin{split}
        {\boldsymbol{\mathbf{y}}}&=\alpha{\boldsymbol{\iota}}+\rho{\boldsymbol{\mathbf{W}}}{\boldsymbol{\mathbf{y}}}+{\boldsymbol{\mathbf{X}}}{\boldsymbol{\beta}}+ {\boldsymbol{\mathbf{u}}},\\
        {\boldsymbol{\mathbf{u}}}&=\lambda{\boldsymbol{\mathbf{W}}}{\boldsymbol{\mathbf{u}}}+{\boldsymbol{\varepsilon}}
\end{split}
\end{equation}

The SAC specification has demonstrated a rather poor performance in
Monte Carlo simulations \citep{Ruttenauer.2022a}. Moreover, it has been
argued that the SAC specification has severe theoretical drawbacks in
applied research, and that its popularity (among econometricians) mainly
stems from the fact that it constitutes an interesting estimation
problem \citep{LeSage.2014}.

\hypertarget{general-nesting-spatial-model-gns}{%
\subsection{General Nesting Spatial Model
(GNS)}\label{general-nesting-spatial-model-gns}}

Finally, the General Nesting Spatial Model (GNS) nests all three
processes: spatial interdependence, clustering on unobservables, and
spillovers in covariates. It can be written as a full combination of
SAR, SEM, and SLX:

\begin{equation}
\begin{split}
        {\boldsymbol{\mathbf{y}}}&=\alpha{\boldsymbol{\iota}}+\rho{\boldsymbol{\mathbf{W}}}{\boldsymbol{\mathbf{y}}}+{\boldsymbol{\mathbf{X}}}{\boldsymbol{\beta}}+{\boldsymbol{\mathbf{W}}}{\boldsymbol{\mathbf{X}}}{\boldsymbol{\theta}}+ {\boldsymbol{\mathbf{u}}},\\
        {\boldsymbol{\mathbf{u}}}&=\lambda{\boldsymbol{\mathbf{W}}}{\boldsymbol{\mathbf{u}}}+{\boldsymbol{\varepsilon}}
\end{split}
\end{equation}

One could be inclined to think that the General Nesting Spatial Model is
superior compared to the more restricted models with two or one source
of spatial dependence. However, in practice the GNS is rather useless as
an estimation model, as it is only weakly identifiable at best
\citep{Gibbons.2012}. This is analogous to Manski's reflection problem
on neighbourhood effects \citep{Manski.1993}: if people in the same
group behave similarly, this can be because a) imitating behaviour of
the group (\(\boldsymbol{\mathbf{W}} \boldsymbol{\mathbf{Y}}\)), b)
members of the same group are exposed to the same external circumstances
(\(\boldsymbol{\mathbf{W}} \boldsymbol{\varepsilon}\)), and c) exogenous
characteristics of the group members
(\(\boldsymbol{\mathbf{W}} \boldsymbol{\mathbf{X}}\)) influence the
behaviour. \emph{We just cannot separate those in observational data.}

All of the models above assume different data generating processes (DGP)
leading to the observed spatial pattern. Although there are
specifications tests, it is generally not possible to let the data
decide which one is the true underlying DGP
\citep{Cook.2020, Ruttenauer.2022a}. There may however be theoretical
reasons to guide the model specification \citep{Cook.2020}. SAR is the
most commonly used model, but it is definitely not the best choice in
many applications. Various studies
\citep{HalleckVega.2015, Ruttenauer.2022a, Wimpy.2021} highlight the
advantages of the relative simple SLX model. Moreover, this
specification can be incorporated in any other statistical method, such
as non-linear estimators or machine learning algorithms.

Note that missing values create a problem in spatial data analysis. For
instance, in a local spillover model with an average of 10 neighbours,
two initial observations with missing values will lead to 20 missing
values in the spatially lagged variable. For global spillover models,
one initial observation with missing values is connect to many other
observations as a higher order neighbour, and thus creates an excess
amount of missings. Depending on the data, units with missings can
either be dropped and omitted from the initial weights creation, or we
need to impute the data first, e.g.~using interpolation or Kriging.
Similarly, islands (i.e units without neighbours) create problems in the
estimation procedure. If this is a very small number of observations,
they can be dropped. Otherwise, distance or k-nearest neighbours may be
alternative options for \(\boldsymbol{\mathbf{W}}\) that circumvent this
problem.

\hypertarget{spatial-impacts}{%
\section{Spatial Impacts}\label{spatial-impacts}}

As shown in Figure Figure~\ref{fig-models}, models withe a SAR-like
process have a feedback loop in the outcome: if my \(x\) influences my
\(y\), this change in my \(y\) will influence my neighbour's \(y\),
which will influence their neighbours' \(y\) and also my own \(y\) again
(I am second order neighbour of my neighbour). We thus cannot interpret
coefficients as marginal or partial effects in SAR, SAC, and SDM
\citep{Anselin.2003, LeSage.2009, Kelejian.2017}. This is similar to
auto-regressive time-series models where we have long-term effects due
to a one unit change in \(x_{it}\). We thus differentiate between the
effects in SAR-like models and those in models without an
auto-regressive (endogenous) outcome term: while SAR, SAC, and SDM
assume \textbf{global} spatial dependence, SLX and SDEM assume
\textbf{local} spatial dependence
\citep{Anselin.2003, HalleckVega.2015, LeSage.2009}.\footnote{Note that
  SEM assumes no spatial effects, as all the spatial dependences comes
  from nuisance} Consequently, also interpretation of the coefficients
differs between models with endogenous feedback loops and those with
only local spillovers.

\hypertarget{global-spillovers}{%
\subsection{Global spillovers}\label{global-spillovers}}

To see the meaning of marginal effects in SAR-like models, we have to
consider its reduced form:

\begin{equation} \label{eq:sarred}
\begin{split}
  {\boldsymbol{\mathbf{y}}} &=({\boldsymbol{\mathbf{I}}_N}-\rho {\boldsymbol{\mathbf{W}}})^{-1}({\boldsymbol{\mathbf{X}}}{\boldsymbol{\beta}}+ {\boldsymbol{\varepsilon}}),
\end{split}
\end{equation}

where \({\boldsymbol{\mathbf{I}}_N}\) is an \(N \times N\) diagonal
matrix (diagonal elements equal 1, 0 otherwise). If interpreting
regression results, we are usually interested in marginal or partial
effects (the association between a unit change in \(X\) and \(Y\)). We
obtain these effects by looking at the first derivative. When taking the
first derivative of the explanatory variable
\({\boldsymbol{\mathbf{x}}}_k\) from the reduced form in
(\ref{eq:sarred}) to interpret the partial effect of a unit change in
variable \({\boldsymbol{\mathbf{x}}}_k\) on
\({\boldsymbol{\mathbf{y}}}\), we receive

\[
\frac{\partial {\boldsymbol{\mathbf{y}}}}{\partial {\boldsymbol{\mathbf{x}}}_k}=\underbrace{({\boldsymbol{\mathbf{I}}_N}-\rho {\boldsymbol{\mathbf{W}}})^{-1}}_{N \times N}\beta_k,
\]

for each covariate \(k=\{1,2,...,K\}\). The partial derivative with
respect to \({\boldsymbol{\mathbf{x}}}_k\) produces an \(N \times N\)
matrix, thereby representing the partial effect of each unit \(i\) onto
the focal unit \(i\) itself and all other units
\(j=\{1,2,...,i-1,i+1,...,N\}\). The \(N \times N\) dimensional term
\(({\boldsymbol{\mathbf{I}}_N}-\rho {\boldsymbol{\mathbf{W}}})^{-1}\) is
also called spatial multiplier matrix. Intuitively, this multiplier
matrix equals a power series:\footnote{A power series of
  \(\sum\nolimits_{k=0}^\infty {\boldsymbol{\mathbf{W}}}^k\) converges
  to \(({\boldsymbol{\mathbf{I}}}-{\boldsymbol{\mathbf{W}}})^{-1}\) if
  the maximum absolute eigenvalue of \({\boldsymbol{\mathbf{W}}} < 1\),
  which is ensured by standardizing \({\boldsymbol{\mathbf{W}}}\).}

\begin{equation} 
\begin{split}
  ({\boldsymbol{\mathbf{I}}_N}-\rho {\boldsymbol{\mathbf{W}}})^{-1}\beta_k 
  =({\boldsymbol{\mathbf{I}}_N} + \rho{\boldsymbol{\mathbf{W}}} + \rho^2{\boldsymbol{\mathbf{W}}}^2 + \rho^3{\boldsymbol{\mathbf{W}}}^3 +   ...)\beta_k 
  = ({\boldsymbol{\mathbf{I}}_N} + \sum_{h=1}^\infty \rho^h{\boldsymbol{\mathbf{W}}}^h)\beta_k ,
\end{split}
\end{equation}

where the identity matrix contains the direct effects and the sum over
\(\rho^h{\boldsymbol{\mathbf{W}}}^h\) represents the first and higher
order indirect effects, including the feedback loops. It implies that a
change in one unit \(i\) does not only affect the direct neighbours but
passes through the whole system towards higher-order neighbours, where
the impact declines with distance within the neighbouring system. Global
indirect impacts thus are `multiplied' by influencing a) direct
neighbours as specified in \(\boldsymbol{\mathbf{W}}\) and b) indirect
neighbours not connected according to \(\boldsymbol{\mathbf{W}}\), with
c) additional feedback loops between those neighbours.

Consider a minimal example with 5 observations, and assume the weights
matrix \(\tilde{\boldsymbol{\mathbf{W}}}\) and its row-normalised
version \(\boldsymbol{\mathbf{W}}\) look as follows:

\begin{equation} 
\begin{split}
\tilde{\boldsymbol{\mathbf{W}}} = \begin{pmatrix}
      0 & 1 & 0 & 1 & 0 \\
      1 & 0 & 1 & 0 & 1 \\
      0 & 1 & 0 & 1 & 0 \\
      1 & 0 & 1 & 0 & 1 \\
      0 & 1 & 0 & 1 & 0
      \end{pmatrix}, ~
\boldsymbol{\mathbf{W}} = \begin{pmatrix}
      0 & 0.5 & 0 & 0.5 & 0 \\
      0.33 & 0 & 0.33 & 0 & 0.33 \\
      0 & 0.5 & 0 & 0.5 & 0 \\
      0.33 & 0 & 0.33 & 0 & 0.33 \\
      0 & 0.5 & 0 & 0.5 & 0
      \end{pmatrix}.      
\end{split}
\end{equation}

Assume that we have relatively strong spatial interdependence with
\(\rho = 0.6\). If we want to get the total effect of \(X\) on \(Y\), we
need to combine the direct effects on the diagonal and the indirect
effects on the off-diagonal.

\begin{equation} 
\begin{split}
\underbrace{\boldsymbol{\mathbf{I}}_N}_{N \times N} - \underbrace{\rho \boldsymbol{\mathbf{W}}}_{N \times N} &=
\begin{pmatrix}
      1 & 0 & 0 & 0 & 0 \\
      0 & 1 & 0 & 0 & 0 \\
      0 & 0 & 1 & 0 & 0 \\
      0 & 0 & 0 & 1 & 0 \\
      0 & 0 & 0 & 0 & 1
      \end{pmatrix} - 
\begin{pmatrix}
      0 & 0.3 & 0 & 0.3 & 0 \\
      0.2 & 0 & 0.2 & 0 & 0.2 \\
      0 & 0.3 & 0 & 0.3 & 0 \\
      0.2 & 0 & 0.2 & 0 & 0.2 \\
      0 & 0.3 & 0 & 0.3 & 0
      \end{pmatrix}\\
& = \begin{pmatrix}
      1 & -0.3 & 0 & -0.3 & 0 \\
      -0.2 & 1 & -0.2 & 0 & -0.2 \\
      0 & 0.3 & 1 & 0.3 & 0 \\
      -0.2 & 0 & -0.2 & 1 & -0.2 \\
      0 & -0.3 & 0 & -0.3 & 1
      \end{pmatrix}.
\end{split}
\end{equation}

Finally, we take the inverse and calculate the spatial multiplier matrix

\begin{equation} 
\begin{split}
\underbrace{(\boldsymbol{\mathbf{I}}_N - \rho \boldsymbol{\mathbf{W}})^{-1}}_{N \times N} &=
\begin{pmatrix}
      1.1875 & 0.46875 & 0.1875 & 0.46875 & 0.1875 \\
      0.3125 & 1.28125 & 0.3125 & 0.28125 & 0.3125 \\
      0.1875 & 0.46875 & 1.1875 & 0.46875 & 0.1875 \\
      0.3125 & 0.28125 & 0.3125 & 1.28125 & 0.3125 \\
      0.1875 & 0.46875 & 0.1875 & 0.46875 & 1.1875
      \end{pmatrix}.
\end{split}
\end{equation}

The \(N \times N\) multiplier matrix
\((\boldsymbol{\mathbf{I}}_N - \rho \boldsymbol{\mathbf{W}})^{-1}\) has
diagonal elements \(>1\): these include direct effects and also feedback
loops, which amplify the direct impact: my \(x\) influences my \(y\)
directly, but my \(y\) then influences my neighbour's \(y\), which then
influences my \(y\) again (and other neighbour's \(y\)s). The influence
of my \(x\) on my \(y\) includes a spatial multiplier effect. To get the
partial effect of a change in \(\boldsymbol{\mathbf{x}}_1\), we need to
multiply the coefficient estimate \(\hat\beta_1\) from the SAR model
with the spatial multiplier matrix. Assume we have
\(\hat\beta_1 = 0.1\), then the partial effect is given by
\(N \times N\) matrix

\begin{equation} 
\begin{split}
\frac{\partial {\boldsymbol{\mathbf{y}}}}{\partial {\boldsymbol{\mathbf{x}}}_1}=\underbrace{({\boldsymbol{\mathbf{I}}_N}-\rho {\boldsymbol{\mathbf{W}}})^{-1}}_{N \times N}\hat\beta_1 = 
\begin{pmatrix}
      0.11875 & 0.046875 & 0.01875 & 0.046875 & 0.01875 \\
      0.03125 & 0.128125 & 0.03125 & 0.028125 & 0.03125 \\
      0.01875 & 0.046875 & 0.11875 & 0.046875 & 0.01875 \\
      0.03125 & 0.028125 & 0.03125 & 0.128125 & 0.03125 \\
      0.01875 & 0.046875 & 0.01875 & 0.046875 & 0.11875
      \end{pmatrix} = \boldsymbol{\Omega}.
\end{split}
\end{equation}

The partial effects matrix \(\boldsymbol{\Omega}\) contains the effect
of each unit \(i\) on itself on the diagonal (including feedback loops)
and the effect on each other unit \(j\) on the off-diagonal. In theory,
\(m_{31} = 0.01875\) tells us that a one-unit change of \(x_1\) in
observation 1 correlates with a \(0.01875\) unit change in the outcome
of observation 3. The \(i\)th row of the matrix \(\boldsymbol{\Omega}\)
represent the impacts \textbf{on} individual observation \(i\), whereas
the \(j\)th column contains the impacts \textbf{from} an individual
observation \(j\) \citep{Anselin.2003, LeSage.2009, LeSage.2014}.
However, the variation across these individual effects depends foremost
on the weights matrix \(\boldsymbol{\mathbf{W}}\). They are not
individual estimates, and it is advisable to not interpret these
individual effects, but rather refer to their summary measures (see
below).

Substantively Interpreting these global spillover effects can be a bit
tricky. The global spillover effects can be understood as a diffusion
process. For example, an exogenous event may increase the house prices
in one district of a city, thus leading to an adaptation of house prices
in neighbouring districts, which then leads to further adaptations in
other units (the neighbours of the neighbours), thereby globally
diffusing the effect of the exogenous event due to the endogenous lag of
\(\boldsymbol{\mathbf{y}}\) term. Yet, those processes happen over time.
In a cross-sectional framework, \citet{Anselin.2003} proposes an
interpretation as an equilibrium outcome, where the partial impact
represents an estimate of how this long-run equilibrium would change due
to a change in \({\boldsymbol{\mathbf{x}}}_k\) \citep{LeSage.2014}.

\hypertarget{local-spillovers}{%
\subsection{Local spillovers}\label{local-spillovers}}

In contrast, the spatial spillover effects of SLX and SDEM are local
spillover effects. They can be interpreted as the effect of a one unit
change of \({\boldsymbol{\mathbf{x}}}_k\) in the spatially weighted
neighbouring observations on the dependent variable of the focal unit.
It is the effect of the weighted average value among neighbours. When
using a row-normalised contiguity weights matrix,
\({\boldsymbol{\mathbf{W}}} {\boldsymbol{\mathbf{x}}}_k\) is the simple
mean of \({\boldsymbol{\mathbf{x}}}_k\) in the neighbouring units.

Assume we have \(k =2\) covariates, then

\begin{equation} 
\begin{split}
\underbrace{\boldsymbol{\mathbf{W}}}_{N \times N}  \underbrace{\boldsymbol{\mathbf{X}}}_{N \times 2} \underbrace{\boldsymbol{\theta}}_{2 \times 1} & = 
\begin{pmatrix}
      0 & 0.5 & 0 & 0.5 & 0 \\
      0.33 & 0 & 0.33 & 0 & 0.33 \\
      0 & 0.5 & 0 & 0.5 & 0 \\
      0.33 & 0 & 0.33 & 0 & 0.33 \\
      0 & 0.5 & 0 & 0.5 & 0
  \end{pmatrix}
  \begin{pmatrix}
      3 & 120 \\
      4 & 140 \\
      1 & 200 \\
      8 & 70  \\
      5 & 250 
  \end{pmatrix}
    \begin{pmatrix}
      \theta_1 \\
      \theta_2 
  \end{pmatrix}\\
 & =   
 \begin{pmatrix}
      6 & 105 \\
      3 & 190 \\
      6 & 105 \\
      3 & 190  \\
      6 & 105 
  \end{pmatrix}
 \begin{pmatrix}
      \theta_1 \\
      \theta_2
  \end{pmatrix}\\
\end{split} 
\end{equation}

Only direct neighbours -- as defined in \({\boldsymbol{\mathbf{W}}}\) --
contribute to those local spillover effects. The
\(\hat{\boldsymbol{\theta}}\) coefficients only estimate how my direct
neighbour's \(\boldsymbol{\mathbf{X}}\) values influence my own outcome
\(\boldsymbol{\mathbf{y}}\). There are no higher order neighbours
involved as long as we do not explicitly specify such higher order
processes, nor are there any feedback loops due to interdependence.

In consequence, local and global spillover effects represent two
distinct kinds of spatial spillover effects \citep{LeSage.2014}. The
interpretation of local spillover effects is straightforward: it is the
effect of a change in \(x_j\) among local neighbours on the outcome of
the focal unit \(y_i\). Global spillover effects are a bit more
complicated: it is the effect that a change in one unit \(x_j\) has on
the entire system of neighbours, bringing \(\boldsymbol{\mathbf{y}}\) on
a new equilibrium outcome.

\hypertarget{summary-measures}{%
\subsection{Summary measures}\label{summary-measures}}

Marginal or partial effects in SAR-like models are given by an
\(N \times N\) matrix of effects. However, since reporting the
individual partial effects is usually not of interest,
\citet{LeSage.2009} proposed to average over these effect matrices.
While the average diagonal elements of the effects matrix
\((\boldsymbol{\mathbf{I}}_N - \rho \boldsymbol{\mathbf{W}})^{-1}\)
represent the so called direct impacts of variable
\({\boldsymbol{\mathbf{x}}}_k\), the average column-sums of the
off-diagonal elements represent the so called indirect impacts (or
spatial spillover effects).

\begin{longtable}[]{@{}
  >{\centering\arraybackslash}p{(\columnwidth - 6\tabcolsep) * \real{0.1507}}
  >{\centering\arraybackslash}p{(\columnwidth - 6\tabcolsep) * \real{0.3425}}
  >{\centering\arraybackslash}p{(\columnwidth - 6\tabcolsep) * \real{0.3562}}
  >{\centering\arraybackslash}p{(\columnwidth - 6\tabcolsep) * \real{0.1507}}@{}}
\toprule\noalign{}
\begin{minipage}[b]{\linewidth}\centering
Model
\end{minipage} & \begin{minipage}[b]{\linewidth}\centering
Direct Impacts
\end{minipage} & \begin{minipage}[b]{\linewidth}\centering
Indirect Impacts
\end{minipage} & \begin{minipage}[b]{\linewidth}\centering
type
\end{minipage} \\
\midrule\noalign{}
\endhead
\bottomrule\noalign{}
\endlastfoot
OLS/SEM & \(\beta_k\) & -- & -- \\
SAR/SAC & Diagonal elements of
\(({\boldsymbol{\mathbf{I}}}-\rho{\boldsymbol{\mathbf{W}}})^{-1}\beta_k\)
& Off-diagonal elements of
\(({\boldsymbol{\mathbf{I}}}-\rho{\boldsymbol{\mathbf{W}}})^{-1}\beta_k\)
& global \\
SLX/SDEM & \(\beta_k\) & \(\theta_k\) & local \\
SDM & Diagonal elements of
\(({\boldsymbol{\mathbf{I}}}-\rho{\boldsymbol{\mathbf{W}}})^{-1}\left[\beta_k+{\boldsymbol{\mathbf{W}}}\theta_k\right]\)
& Off-diagonal elements of
\(({\boldsymbol{\mathbf{I}}}-\rho{\boldsymbol{\mathbf{W}}})^{-1}\left[\beta_k+{\boldsymbol{\mathbf{W}}}\theta_k\right]\)
& global \\
\end{longtable}

Note that impacts in \textbf{SAR and SAC are bound to a common ratio}
between direct and indirect impacts. SAR and SAC models only estimate
one single spatial multiplier coefficient. Thus direct and indirect
impacts have a common ratio (say \(\phi\)) across all covariates: if
\(\beta_1^{direct} = \phi\beta_1^{indirect}\), then
\(\beta_2^{direct} = \phi\beta_2^{indirect}\),
\(\beta_k^{direct} = \phi\beta_k^{indirect}\). For specifications
including a lagged version of \(\boldsymbol{\mathbf{X}}\), in contrast,
we estimate a local spatial effect for each unique covariate, plus an
additional spatial multiplier in case of an SDM. SLX-like specification
are thus much more flexible. Usually, impact measures come with
simulation based inferential statistics \citep{Bivand.2015}.

\hypertarget{model-selection}{%
\section{Model selection}\label{model-selection}}

Various spatial model specifications can be used to account for the
spatial structure of the data. Selecting the correct model specification
remains a crucial task in applied research. There are two empirical
strategies for model selection: a specific-to-general or a
general-to-specific approach \citep{Florax.2003, Mur.2009}. However,
both come with severe drawbacks.

The specific-to-general approach is more common in spatial econometrics.
This approach starts with the most basic non-spatial model and tests for
possible misspecification due to omitted autocorrelation in the error
term or the dependent variable. \citet{Anselin.1996} has proposed to use
Lagrange multiplier (LM) tests for the hypotheses \(H_0\): \(\lambda=0\)
and \(H_0\): \(\rho=0\), which are robust against the alternative source
of spatial dependence. The specific-to-general approach based on the
robust LM test offers a good performance in distinguishing between SAR,
SEM, and non-spatial OLS \citep{Florax.2003}. Still, the test disregards
the presence of spatial dependence from local spillover effects
(\(\theta\) is assumed to be zero), as resulting from an SLX-like
process. \citet{Cook.2020} show theoretically that an SLX-like
dependence structure leads to the rejection of both hypotheses \(H_0\):
\(\lambda=0\) and \(H_0\): \(\rho=0\), although no autocorrelation is
present \citep{Elhorst.2017, Ruttenauer.2022a}.

The general-to-specific approach follows the opposite direction. It
starts with the most general model and stepwise imposes restrictions on
the parameters of this general model. In theory, we would 1) start with
a GNS specification and 2) subsequently restrict the model to simplified
specifications based on the significance of parameters in the GNS
\citep{HalleckVega.2015}. The problem with this strategy is that the GNS
is only weakly identified and, thus, is of little help in selecting the
correct restrictions \citep{Burridge.2016}. The most intuitive
alternative would be to start with one of the two-source models SDM,
SDEM, or SAC. This, however, bears the risk of imposing the wrong
restriction in the first place \citep{Cook.2020}. Furthermore,
\citet{Cook.2020} show that more complicated restrictions are necessary
to derive all single-source models from SDEM or SAC specifications.

Some argue that the best way of choosing the appropriate model
specification is to exclude one or more sources of spatial dependence --
autocorrelation in the dependent variable, autocorrelation in the
disturbances, or spatial spillover effects of the covariates -- by
design \citep[\citet{Gibbons.2015}]{Gibbons.2012}. Natural experiments
would be the best way of making one or more sources of spatial
dependence unlikely, thereby restricting the model alternatives to a
subset of all available models. However, the opportunities to use
natural experiments are restricted in social sciences, making it a
favourable but often impractical way of model selection.
\citet{Cook.2020} and \citet{Ruttenauer.2022a} argue that theoretical
considerations should guide the model selection. 1) Rule out some
sources of spatial dependence by theory, and thus restrict the
specifications to a subset, and 2) theoretical mechanisms may guide the
choice of either global or local spillover effects.

A recent simulation study \citep{Ruttenauer.2022a} has shown that SLX,
SDM, and SDEM are preferable if all sources of dependence may be
present. Besides that, the SLX is the most simple specification, as it
can easily be estimated by OLS. Given that
\(\boldsymbol{\mathbf{W}} \boldsymbol{\mathbf{X}}\) is just another
variable, SLX can easily be combined with non-linear models or other
more complicated model specifications, such as panel estimators or
machine learning algorithms. Similar conclusions are supported by
\citet{Wimpy.2021}, and also Jeffrey Wooldridge argued for SLX as the
only reasonable spatial specification in a Tweet from 2021 called ``I
will use spatial lags of X, not spatial lags of Y'' \footnote{Tweet on
  using SLX by J. Wooldridge on Twitter:
  \url{https://twitter.com/jmwooldridge/status/1369460526770753537}}.

\hypertarget{house-prices-in-london}{%
\section{House prices in London}\label{house-prices-in-london}}

As an example to compare the different spatial model specifications, we
estimate the effect of local characteristics such as green space and
public transport connectivity on the median house price. The relation
between environmental characteristics and housing choice and prices has
been investigated in several studies
\citep{Anselin.2008, Kley.2021, Liebe.2023}. The data for the current
example was retrieved from the London Datastore\footnote{For house
  prices, see:
  \url{https://data.london.gov.uk/dataset/average-house-prices}. For
  London accessibility scores see:
  \url{https://data.london.gov.uk/dataset/public-transport-accessibility-levels}},
the 2011 Census\footnote{For UK demographics, see:
  \url{https://www.nomisweb.co.uk/sources/census_2011}} and
OpenStreetMaps and combined at the Middle Layer Super Output Areas
(MSOA). There are 983 MSOAs in London with an average population size of
around 8,000 residents. The script for compiling and preparing the data
can be found in the Supplementary Materials. All data preparation and
analysis were performed with the statistical software R. For a
comprehensive overview of spatial software see \citet{Bivand.2021} or
\citet{Pebesma.2023}.

\begin{figure}

{\centering \includegraphics{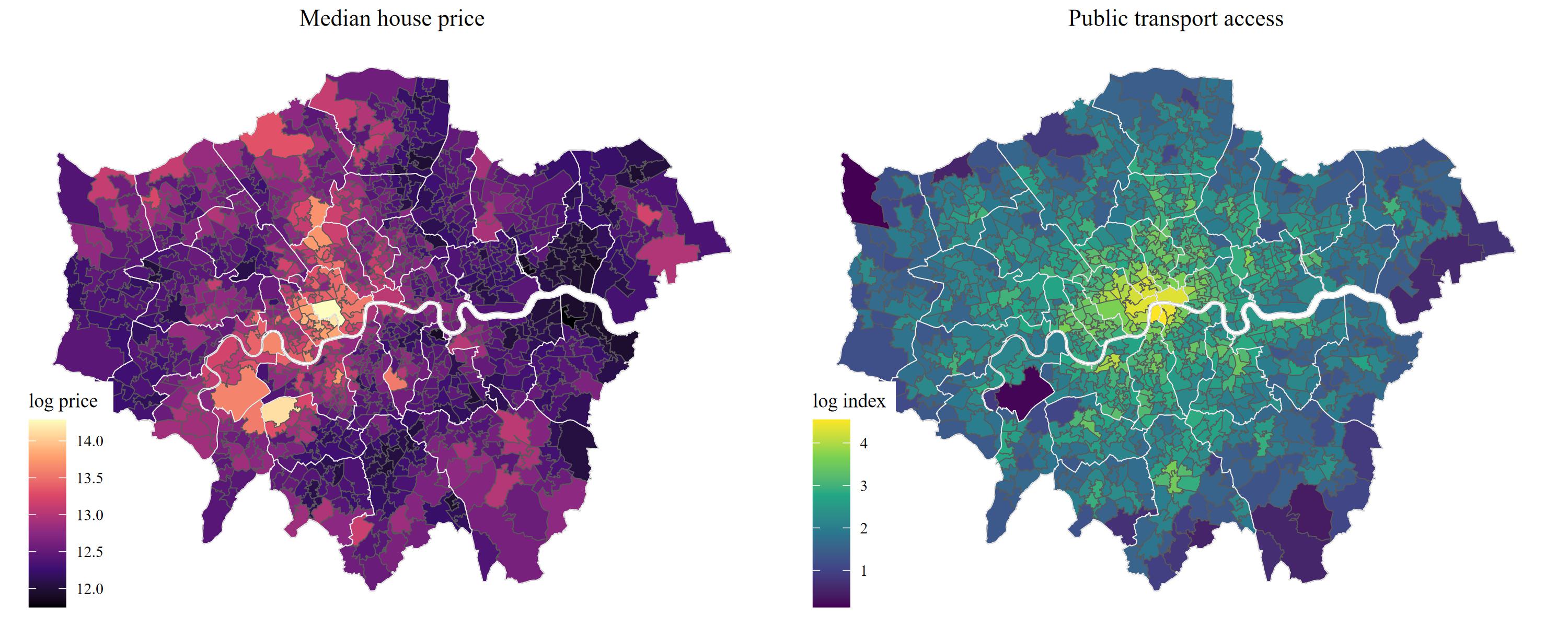}

}

\caption{\label{fig-map}Spatial distribution of log-transformed median
house prices and transport accessibility across London.}

\end{figure}

Figure Figure~\ref{fig-map} shows an unclassified choropleth map of
house prices and public transport access across London, both log-scaled
for mapping. As we would expect, both indicators follow a relatively
strong spatial of positive autocorrelation: house prices first decrease
with increasing distance to the centre, and then seem to increase again
in suburban areas. Moreover, there seems to be a pattern of higher
prices towards the west and particularly high prices around Hyde Park.
Public transport accessibility steadily decreases with distance to the
city centre. Spatial regression models thus seem to be important here
for two reasons: a) observations are not independent of each other but
follow clear spatial patterns, and b) surrounding / adjacent urban
characteristics likely play a role for housing demand and prices in the
focal unit as well.

In Table 1, we regress the median house price in 2011 on the area (in
km\^{}2) covered by green space according to OpenStreetMaps, an index of
public transport access (ranging from 0-low accessibility to 100-high
accessibility), and several population characteristics from the 2011
census such as population density, the percent of non-UK residents and
the percent of social housing. Reported are results form (1) non-spatial
OLS, (2) Spatial Autoregressive (SAR), (3) Spatial Error Model (SEM),
(4) Spatial Lag of X (SLX), (5) Spatial Durbin Model (SDM), (6) and
Spatial Durbin Error Model (SDEM). All variables were standardized
before estimation, and we thus interpret coefficients in standard
deviations. Note that we do not estimate results for Spatial
Autoregressive Combined (SAC) models because of its severe drawbacks for
applied research \citep{LeSage.2014}.

\begin{table}
\caption{Spatial regression models. Outcome variable: median house price.}
\begin{center}
\begin{scriptsize}
\begin{tabular}{l D{.}{.}{4.6} D{.}{.}{4.6} D{.}{.}{4.6} D{.}{.}{4.6} D{.}{.}{4.6} D{.}{.}{4.6}}
\hline
 & \multicolumn{1}{c}{OLS} & \multicolumn{1}{c}{SAR} & \multicolumn{1}{c}{SEM} & \multicolumn{1}{c}{SLX} & \multicolumn{1}{c}{SDM} & \multicolumn{1}{c}{SDEM} \\
\hline
(Intercept)               & 0.000        & -0.012       & 0.022        & 0.007        & 0.002        & 0.007        \\
                          & (0.027)      & (0.017)      & (0.139)      & (0.024)      & (0.015)      & (0.103)      \\
Green space               & 0.204^{***}  & 0.133^{***}  & 0.100^{***}  & 0.136^{***}  & 0.106^{***}  & 0.111^{***}  \\
                          & (0.029)      & (0.018)      & (0.015)      & (0.026)      & (0.016)      & (0.020)      \\
Public transport access   & 0.366^{***}  & 0.097^{***}  & -0.054       & -0.152^{**}  & -0.100^{**}  & -0.042       \\
                          & (0.033)      & (0.021)      & (0.033)      & (0.054)      & (0.034)      & (0.033)      \\
Population density        & 0.189^{***}  & 0.055^{*}    & -0.094^{***} & -0.112^{*}   & -0.111^{***} & -0.099^{***} \\
                          & (0.037)      & (0.023)      & (0.027)      & (0.044)      & (0.028)      & (0.029)      \\
Percent non-UK            & -0.033       & -0.050^{*}   & -0.250^{***} & -0.235^{***} & -0.262^{***} & -0.232^{***} \\
                          & (0.033)      & (0.020)      & (0.033)      & (0.053)      & (0.033)      & (0.032)      \\
Percent social housing    & -0.402^{***} & -0.202^{***} & -0.260^{***} & -0.306^{***} & -0.266^{***} & -0.282^{***} \\
                          & (0.032)      & (0.020)      & (0.022)      & (0.035)      & (0.022)      & (0.024)      \\
W Green space             &              &              &              & 0.249^{***}  & -0.029       & 0.050        \\
                          &              &              &              & (0.040)      & (0.026)      & (0.043)      \\
W Public transport access &              &              &              & 0.696^{***}  & 0.239^{***}  & 0.273^{***}  \\
                          &              &              &              & (0.069)      & (0.045)      & (0.073)      \\
W Population density      &              &              &              & 0.455^{***}  & 0.136^{***}  & -0.043       \\
                          &              &              &              & (0.065)      & (0.041)      & (0.075)      \\
W Percent non-UK          &              &              &              & 0.304^{***}  & 0.300^{***}  & 0.242^{***}  \\
                          &              &              &              & (0.066)      & (0.042)      & (0.070)      \\
W Percent social housing  &              &              &              & -0.352^{***} & 0.119^{***}  & -0.135^{*}   \\
                          &              &              &              & (0.053)      & (0.035)      & (0.063)      \\
\hline
Num. obs.                 & 983          & 983          & 983          & 983          & 983          & 983          \\
R$^2$                     & 0.263        &              &              & 0.439        &              &              \\
Adj. R$^2$                & 0.259        &              &              & 0.433        &              &              \\
LR test: statistic        &              & 789.480      & 934.291      &              & 732.684      & 695.097      \\
LR test: p-value          &              & 0.000        & 0.000        &              & 0.000        & 0.000        \\
AIC                       & 2502.492     & 1715.012     & 1570.201     & 2244.135     & 1513.451     & 1551.038     \\
\hline
\multicolumn{7}{l}{\tiny{$^{***}p<0.001$; $^{**}p<0.01$; $^{*}p<0.05$}}
\end{tabular}
\end{scriptsize}
\label{table:coefficients}
\end{center}
\end{table}

Compared to results from conventional non-spatial models, Table 1 comes
with several additions: First, variables starting with a ``W'' (or
``lag'') indicate the spatially lagged variable or in the case of
row-normalized weights matrices the average value of the respective
variable across the local neighbours. Moreover, there are two
auto-regressive parameters: ``rho'' for the estimated auto-correlation
in the dependent variable and ``lambda'' for the estimated
auto-correlation in the error term. In case of the SAR, a highly
significant \(\hat\rho\) coefficient of 0.786 indicates strong positive
spatial auto-correlation in the median house price: the house price in
adjacent areas positively impacts the focal house prices. A
\(\hat\lambda\) of 0.89 in the SEM however indicates that there is very
strong spatial auto-correlation among the (remaining) error variance.
The likelihood ratio test in the goodness-of-fit statistics are highly
significant in both cases, rejecting the NULL of no spatial
auto-correlation.

\begin{figure}

{\centering \includegraphics{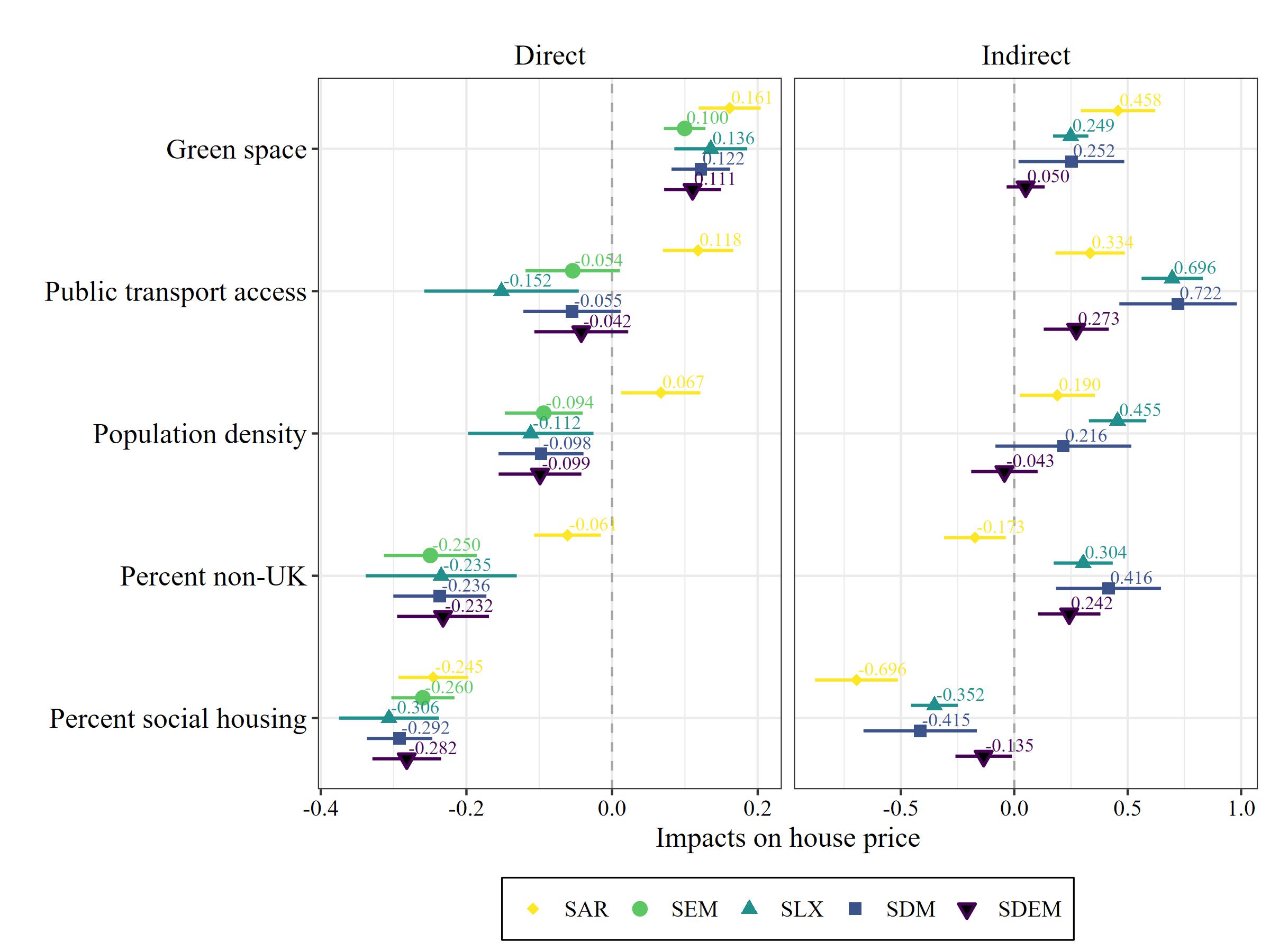}

}

\caption{\label{fig-coefs}Direct and indirect impacts from spatial
regression models. Dependent variable: mean house prices. All vairables
are standardised.}

\end{figure}

Given the strong positive auto-correlation in the dependent variable in
SAR and SDM, we cannot directly interpret the coefficients as marginal
effects. Similar to auto-regressive temporal models, we need to account
for the spatial multiplier effect. For SEM, SLX and SDEM, we could
directly interpret the coefficients of Table 1. However, we plot the
impacts of all five models in Figure Figure~\ref{fig-coefs} for reasons
of comparison. Note that SEM only has direct and no indirect impacts.

We start with the results of the SAR model in Figure
Figure~\ref{fig-coefs}. A one standard-deviation increase of green space
in the focal unit is associated with a 0.161 standard deviation increase
in house prices within the same spatial unit. However, there are also
highly significant diffusion processes. This increase in green space in
the focal unit will also increase house prices in neighbouring units and
the neighbours of these neighbours. This indirect impact will add up to
a 0.458 standard deviation increase in house prices across neighbouring
units connected through the spatial weights system. Similarly, an
increase in public transport accessibility is associated with a 0.118
standard-deviation higher median house price in the unit itself and an
additional 0.334 deviation increase diffusing though the neighbouring
regions. Note that direct and indirect effects are bound to a common
ration, as SAR only estimates one single spatial parameter \(\hat\rho\).
In our case, every indirect impact equals approximately 2.83 times the
direct impact. This is a very restrictive conditions and a severe
drawback of the SAR model.

The SLX - similar to SAR - estimates a positive impact of green space in
the focal and also in adjacent neighbourhoods on house prices in the
focal unit. A one standard deviation in the focal unit is associated
with 0.136 standard-deviations higher house price in the focal unit. If
green spaces in adjacent neighbourhoods increase on average by one
standard deviation, this would increase house prices in the focal unit
by 0.249 standard deviations. Note that the SLX tells a different story
about the effect of public transport access than SAR: there is a
negative direct and a very strong and positive indirect effect. A one
standard deviation increase in public transport access in the focal unit
is associated with -0.152 standard deviations lower house prices. In
contrast, more public transport in the local surrounding (the average
neighbours) is associated with 0.696 standard deviations higher prices.
This is in line with the idea that public transport facilities are
usually not particularly attractive: it is good to have them close but
not too close. The same is true for population density: it is good to
live in a broader area with high population density as indicated by the
indirect impacts (likely indicating high centrality), but the local
neighbourhood should have a low population density as indicated by the
negative direct impact.

We could go further with the other models. However, interpretation in
SDM follows the same logic as SAR, and interpretation in SDEM aligns to
SLX. Interpretation in SEM is analogous to non-spatial OLS, as there are
no indirect impacts. Moreover, it is important to keep in mind that the
indirect impacts are summary measures which sum over all impacts from or
onto neighbouring regions. The indirect public transport effect of 0.696
in SLX would occur if the average public transport access across
neighbours would increase by one standard deviation. This only occurs if
all neighbours would simultaneously increase public transport access by
one standard deviation.

\hypertarget{conclusions}{%
\section{Conclusions}\label{conclusions}}

Interest in spatial research topics has witnessed a surge within the
social sciences, and it is paralleled by the increasing availability of
geo-referenced data. This growing availability carries immense potential
for delving into the analysis of spatial phenomena, such as spillovers
and diffusions. However, it also presents challenges for statistical
estimators. Notably, utilizing non-spatial techniques with spatial data
results in the loss of valuable information. This chapter offers an
extensive overview of common spatial econometrics models that permit the
explicit testing of spatial relationships. For those keen on exploring
spatial panel data, consider the works of \citet{Elhorst.2014} and
\citet{Cook.2023}. Meanwhile, those intrigued by non-linear spatial
models should delve into \citet{LeSage.2009} and \citet{Franzese.2016}.

In the regression framework of this chapter, spatial dependence can be
integrated as three distinct processes: a) Spatial interdependence in
the outcome, b) Clustering of unobservable factors, and c) Spillovers
originating from covariates. In any practical application, it is crucial
to first contemplate potential theoretical underpinnings for spatial
dependence. In the case outlined above, it is plausible to anticipate
dependence in the outcome, as house prices in adjacent neighbourhoods
directly influence prices in the focal units, given that agents or
home-owners rely on price information from surrounding areas. Clustering
of unobservable factors is also evident; attributes like distance to the
city centre or housing age are spatially clustered and likely exert an
influence on house prices and other covariates, potentially causing an
omitted variable bias. Moreover, there are likely spillover effects from
the covariates, where factors like parks, population density, and public
transport access in surrounding neighbourhoods have a direct impact
across neighbourhood borders. Thus, all the models discussed here are
theoretically plausible.

The choice of the correct model specification is often arbitrary,
especially in cases like house price modelling. It is advisable to steer
clear of the Spatial Autoregressive (SAR) and Autoregressive Conditional
(SAC) models, as they come with drawbacks in applied research as
highlighted by \citet{LeSage.2014} and \citet{Ruttenauer.2022a}. Models
with only one estimated spatial parameter across all covariates, like
SAR and SAC, impose heavy restrictions on indirect impacts, potentially
leading to biased estimates when multiple covariates are involved.
Consequently, it is generally sensible to consider more flexible
specifications such as SLX, Spatial Durbin Model (SDM), and Spatial
Durbin Error Model (SDEM). In our example above, the conclusions derived
from SLX, SDM, and SDEM are fairly consistent, with SDEM being the most
conservative regarding indirect spatial impacts. This aligns with the
model's accounting for spatial clustering among errors, which
encompasses potential confounders. For instance, the indirect impact of
population density diminishes significantly when controlling for the
distance to the city center, explaining why the indirect positive effect
of population density vanishes in SDEM---it is largely confounded by
distance to the city center. Note that controlling for the distance to
the city center would already account for some proportion of the
auto-correlation.

The Spatial Lag Model (SLX) stands out for several reasons: 1) It is
straightforward in its simplicity; 2) Estimation can be performed using
least squares; 3) It can be seamlessly integrated into panel data
models, non-linear models, and machine learning techniques, treating
\(\boldsymbol{\mathbf{W}} \boldsymbol{\mathbf{X}}\) as just another set
of covariates; 4) SLX can be globalised by incorporating higher-order
neighbours such as \(\boldsymbol{\mathbf{W}}^2 \boldsymbol{\mathbf{X}}\)
and so forth, allowing for a broader assessment of spatial impacts.

A topic deserving more attention is the necessity for spatial
econometric models when working with individual-level survey data merged
with geographic context information. Do we need to account for spatial
structure when adding neighbourhood information to survey data? A common
approach involves multi-level models, which address error dependence.
However, this approach assumes that units living very close to each
other but separated by an arbitrary spatial border are independent ----
a strong assumption. An alternative approach is a spatial error model,
which accommodates spatially clustered errors. For instance,
\citet{Diekmann.2023} presents a compelling example in the field of
environmental inequality, where error models seem more plausible since
it is unlikely that randomly sampled survey respondents directly
influence each other (as assumed in SLX and SAR), but very likely that
neighbouring respondents are exposed to similar unobservable factors.
Nevertheless, one may still wish to investigate the influence of context
effects and their spatial patterns. In such cases, SLX-like
specifications for the context appear reasonable, as demonstrated by
\citet{Haussmann.2023}, who employed spatial SLX specifications to
explore the impact of regional deprivation on right-wing votes at
various spatial scales.

For further exploration in spatial data analysis, I recommend
\citet{Pebesma.2023} as an open-science book on Spatial Data Science,
offering a comprehensive overview of handling and processing spatial
data. \citet{LeSage.2009} and \citet{Kelejian.2017} provide
comprehensive introductions to spatial econometrics, complete with the
necessary mathematical foundations. \citet{Ward.2008} offers an
intuitive introduction to spatial regression models, while
\citet{Elhorst.2012}, \citet{HalleckVega.2015}, \citet{LeSage.2014}, and
\citet{Ruttenauer.2022a} present article-length introductions to spatial
econometrics.

\renewcommand\refname{References}
  \bibliography{references.bib}

\end{document}